\documentclass[preprint,showpacs,preprintnumbers,amsmath,amssymb,superscriptaddress,prb]{revtex4}

\usepackage{graphicx}
\usepackage{dcolumn}
\usepackage{bm}
\usepackage[T1]{fontenc} 
\usepackage{color}
\newcommand{\SMO}{SrMnO$_3$}
\newcommand{\SRMOx}{SrRu$_{1-x}$Mn$_x$O$_3$}
\newcommand{\STMO}{SrTi$_{0.95}$Mn$_{0.05}$O$_3$}
\newcommand{\STO}{SrTiO$_3$}
\newcommand{\NSTO}{Nb:SrTiO$_3$}
\newcommand{\SRMO}{SrRu$_{0.95}$Mn$_{0.05}$O$_3$}
\newcommand{\SRO}{SrRuO$_3$}
\newcommand{\IV}{\textit{I}-\textit{V}}
\begin{document}

\preprint{APS/123-QED}

\title{Negative Differential Resistance Induced by Mn Substitution at \\ \SRO/\NSTO \ Schottky Interfaces}

\author{Yasuyuki Hikita}
 \email{hikita@k.u-tokyo.ac.jp}
 \affiliation{Department of Advanced Materials Science, University of Tokyo, Kashiwa, Chiba 277-8561, Japan}
\author{Lena Fitting Kourkoutis}
 \affiliation{School of Applied and Engineering Physics, Cornell University, Ithaca, New York 14853, USA}
\author{Tomofumi Susaki}%
 \altaffiliation{Present address: Materials and Structures Laboratory, Tokyo Institute of Technology, Yokohama, Kanagawa 226-8503, Japan.}
 \affiliation{Department of Advanced Materials Science, University of Tokyo, Kashiwa, Chiba 277-8561, Japan}
\author{David A. Muller}
 \affiliation{School of Applied and Engineering Physics, Cornell University, Ithaca, New York 14853, USA}
\author{Hidenori Takagi}%
 \affiliation{Department of Advanced Materials Science, University of Tokyo, Kashiwa, Chiba 277-8561, Japan}
\author{Harold Y. Hwang}%
 \affiliation{Department of Advanced Materials Science, University of Tokyo, Kashiwa, Chiba 277-8561, Japan}
 \affiliation{Japan Science and Technology Agency, Kawaguchi, Saitama 332-0012, Japan}

\date{\today}

\begin{abstract}
We observed a strong modulation in the current-voltage characteristics of \SRO/\NSTO \ Schottky junctions by Mn substitution in  \SRO, which induces a metal-insulator transition in bulk. The temperature dependence of the junction ideality factor indicates an increased spatial inhomogeneity of the interface potential with substitution. Furthermore, negative differential resistance was observed at low temperatures, indicating the formation of a resonant state by Mn substitution. By spatially varying the position of the Mn dopants across the interface with single unit cell control, we can isolate the origin of this resonant state to the interface \SRO \ layer. These results demonstrate a conceptually different approach to controlling interface states by utilizing the highly sensitive response of conducting perovskites to impurities.
\end{abstract}


\maketitle
\section{\label{sec:Introduction}introduction}
Schottky junctions using perovskite oxides are systems in which several interesting device properties have been reported, including resistance switching \cite{Beck}, magnetic field sensitive diode \cite{Nakagawa}, and as a prototypical structure for enhanced photocarrier doping \cite{Katsu}. One of the important concepts in Schottky junctions is the formation of interface states which strongly influence barrier formation, and in many cases are responsible for nonideal device performance \cite{EHRoderick}. In elemental metal/conventional semiconductor junctions, interface states are formed as consequences of intrinsic surface reconstructions, impurities or defects on the semiconductor surfaces, or alloying by a metal-semiconductor reaction. In many cases,  the barrier formation processes are driven predominantly by the properties of the \textit{semiconductor}, and are usually less dependent on the properties of the metal, with some exceptions such as in the presence of metal-induced-gap states\cite{Heine}. This is due to the large difference in the bonding character of the constituents, namely spatially oriented covalent bonds in semiconductors and spatially uniform metallic bonds in the metal, which can easily result in amorphous or polycrystalline interfaces.
\\\indent
In contrast, the chemical and structural similarities between metallic and semiconducting perovskites enable the epitaxial growth of oxide heterojunctions. Furthermore, the electrical conductivity in metallic perovskites is often achieved by the delicate energy balance between  competing ground states, as seen in high-$T_c$ cuprates or colossal magnetoresistive manganites \cite{Imada}. Upon doping, many perovskite metals transition through a carrier localized state long before completely becoming an insulator with a well-defined gap in the density of states. When interfaces are formed using such disordered metals, a new type of interface state formation can be anticipated, giving us an additional degree of freedom to manipulate the interface electronic structure from the \textit{metal} side.
\\\indent
To explore these ideas, we have investigated controlled impurity substitution in \SRO/\NSTO \ junctions. This is an ideal system to study because the interface is free from a polar discontinuity which can be a significant source of interface states \cite{Nakagawa_b}, and its barrier formation is not dominated by interface states, as evidenced by the good agreement between the experimentally determined Schottky barrier height \cite{Hikita, Minohara} and the Schottky-Mott relation. Additionally, \NSTO \ has been the most thoroughly investigated perovskite semiconductor in junction structures, such as the Schottky barrier height characterization \cite{Hikita} and the studies of non-linear dielectric constant and its effect on band bending \cite{Yamamoto, Susaki_a}. \SRO, in the absence of chemical substitution, is a representative conducting perovskite, with a ferromagnetic transition temperature $T_C$ = 160 K. It exhibits several unusual transport properties, such as the absence of resistivity saturation up to 1000 K \cite{Allen}. The strong divergence of $dR/dT$ as $T \to T_{C}^{+}$ and the weak divergence when $T \to T_{C}^{-}$ raises questions of the applicability of the standard Boltzmann transport picture, leading to the assignment of \SRO \ as a "bad metal" \cite{Klein}. As can be expected from the unusual behavior in its pure form, \SRO \ is highly sensitive to disorder induced by chemical doping such as Ca \cite{Fukunaga}, Zn, Ni, Co \cite{Pi}, Fe \cite{Bansal} or Mn \cite{Pi, Banerjee, Cao_b} doping, all of which induce metal-insulator transitions. In the case of doping with a 3\textit{d} transition metal ion at the Ru site, the Ru-O bonding network is reduced and carrier localization occurs because of the localized nature of the 3\textit{d} orbitals.
\\\indent
Here, we study the effects of using a strongly correlated metal at the junction interface by examining the temperature dependent \IV \ characteristics of Mn-doped \SRO/\NSTO \ Schottky junctions. From the high temperature slope in the semi-logarithmic plot of the \IV \ characteristics, the spatial distribution of the barrier height increased by Mn substitution. At low temperatures, we observed negative differential resistance even for a single layer of Mn substituted \SRO \ metal at the interface. These results indicate that the high sensitivity of strongly correlated metals to impurities can give rise to characteristic interface state formation from imperfectly screened impurity ions.

\section{\label{sec:Experimental}experimental}
\begin{figure}[t]
  \begin{center}
    \includegraphics[clip]{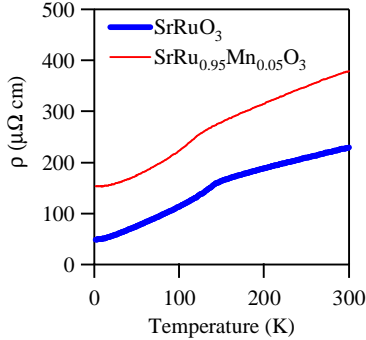}
  \end{center}
  \caption{(Color online) Temperature dependent resistivity ($\rho$) of the \SRO \ (bold) and \SRMO \ (solid) thin films grown on \STO(100) substrates. The increased resistivity and the reduced $T_C$ is evident for the \SRMO \ thin film.}
  \label{fig:Fig1_4.eps}
\end{figure}

Epitaxial thin film structures were fabricated by pulsed laser deposition using a KrF excimer laser with a fluence of 2.0 J/cm$^2$ on \STO(100) \ substrates undoped and Nb = 0.5 wt \% doped in an oxygen partial pressure ($P_{O2}$) of 0.3 Torr (unless otherwise indicated) and at a substrate temperature of 800 $^\circ$C. The deposited film thickness of \SRO \ and \SRMO \ were 800 \AA.  Polycrystalline \SRMOx \ ($x$ = 0, 0.05, 0.15), \SMO, \STMO \ and single crystal \STO \ targets were used.\\\indent
The critical concentration required for the bulk metal-insulator transition in \SRMOx \ is $x$ = 0.4 (Ref. \onlinecite{Cao_b}). At $x$ = 0.05, predominantly studied here, \SRMOx \ is reported to be a metal with a slightly increased resistivity and a reduced $T_C$. The X-ray diffraction patterns of the thin films revealed high quality epitaxial growth and an out-of-plane lattice constant of 3.95 \AA \  for both \SRMO \ and undoped \SRO. As shown in Fig. \ref{fig:Fig1_4.eps}, the temperature dependent resistivity ($\rho$) of the films deposited on insulating \STO(100) substrates exhibited metallic behavior with a $T_C$ and residual resistivity of 150 K and 50 $\mu\Omega$cm for \SRO, and 140 K and 150 $\mu\Omega$cm for \SRMO.
\section{\label{sec:results}results}
\subsection{\label{sec:inhomogeneity}Temperature dependent \IV characteristics}
\begin{figure}[t]
  \begin{center}
    \includegraphics[clip]{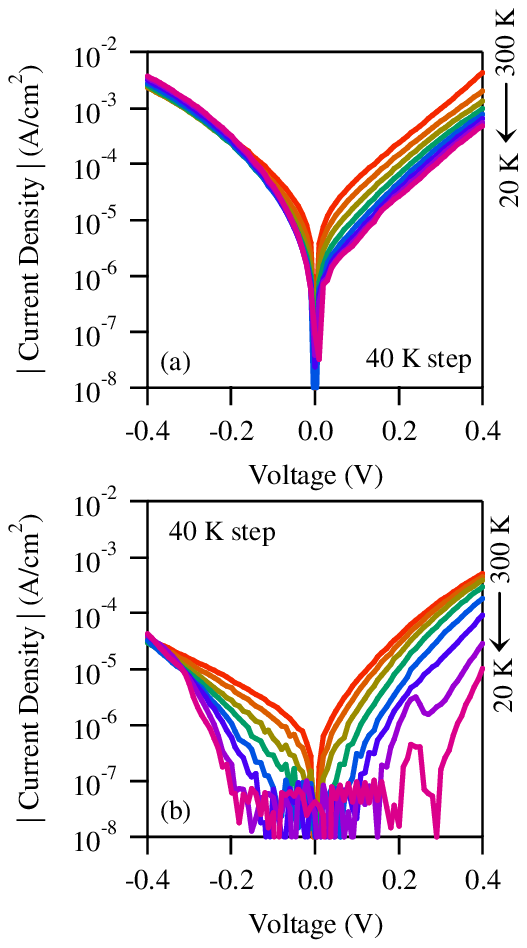}
  \end{center}
  \caption{(Color online) Temperature dependence of the \IV \ characteristics in (a) \SRO/\NSTO \  and (b) \SRMO/\NSTO \ junctions. Measurements taken between 300 K and 20 K in 40 K steps.}
  \label{fig:Fig2_4.eps}
\end{figure}
\begin{figure}[t]
  \begin{center}
    \includegraphics[clip]{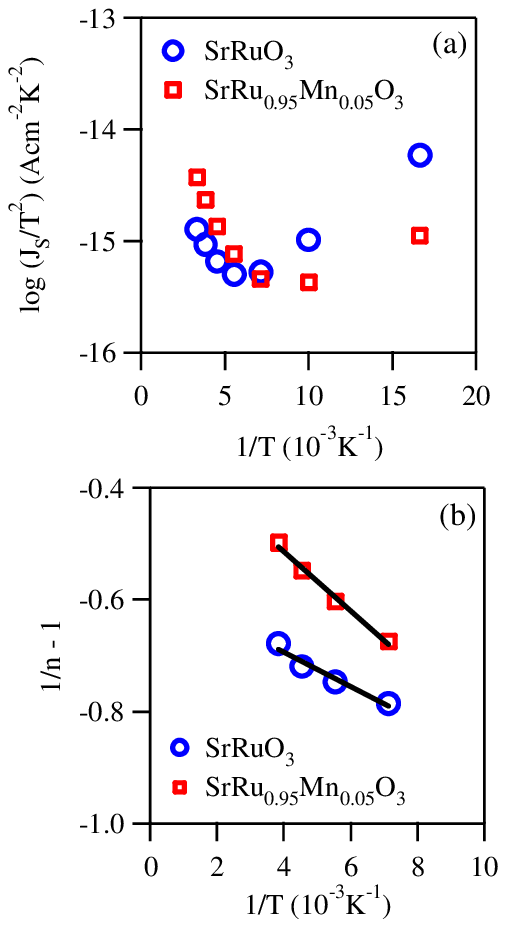}
  \end{center}
  \caption{(Color online) (a) Richardson plot of \SRO \ (\SRMO)/\NSTO \ junctions showing a crossover from thermoionic field emission to field emission. (b) Evaluation of the Schottky barrier height spatial inhomogeneity from the ideality factors. The circles (squares) denote \SRO \ (\SRMO) and the lines are linear fits to the experimental data in the high temperature regime.}
  \label{fig:Fig2a_4.eps}
\end{figure}
\begin{figure}[t]
  \begin{center}
    \includegraphics[clip]{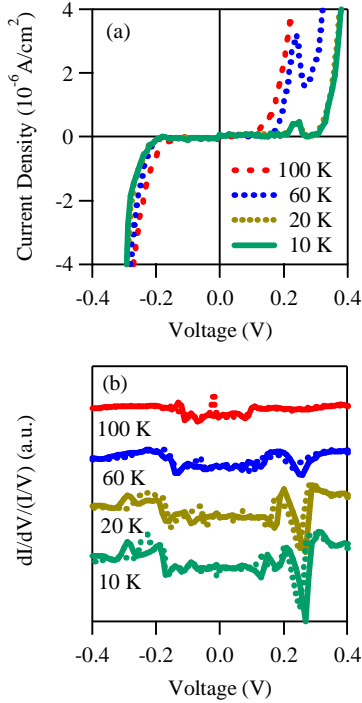}
  \end{center}
  \caption{(Color online) Temperature dependence of the (a) \IV \ characteristics and (b) the normalized conductance in a \SRMO/\STO \ junction. Solid (dotted) lines in (b) denote a voltage sweeps in the positive  (negative) direction. NDR emerges as the temperature is decreased.}
  \label{fig:Fig3_4.eps}
\end{figure}

The temperature dependent \IV \ characteristics for \SRO/\NSTO \ and \SRMO/\NSTO \ are shown in Fig. \ref{fig:Fig2_4.eps}(a) and (b), respectively. Ohmic contacts were made by Ag paste directly on the ruthenates and Al deposition on \NSTO. The polarity of the applied bias is defined as positive voltage applied to the ruthenate.
\\\indent
In the case of the \SRO\ junction, the forward bias current in the semi-logarithmic plot is linearly proportional to the bias voltage with an overall shift to higher voltages at lower temperatures, indicating near ideal Schottky behavior. The current transport mechanism of these junctions was determined from the Richardson plot shown in Fig. \ref{fig:Fig2a_4.eps}(a). The reduced slope at high temperatures indicates the contribution of thermoionic field emission (TFE), which is consistent with the large reverse bias current caused by the high doping concentration in \NSTO. The gradual increase in $J_S/T^2$ at low temperatures corresponds to the dominance of the field emission (FE) process in this temperature regime. Here, $J_S$ is the saturation current density obtained by a linear extrapolation of the forward biased region of the \IV \ characteristics to zero voltage. The temperature dependence of the two junctions is very similar to what has been observed in Au/Nb:SrTiO$_3$ Schottky junctions \cite{Susaki_a, Hasegawa}. \\\indent
In contrast, two apparent differences can be seen in the \IV \ characteristics of the \SRMO/\NSTO \ junction, compared to the \SRO/\NSTO \ junction (Fig. \ref{fig:Fig2_4.eps}). First, the large decrease in the current over the measured voltage range by almost an order of magnitude, and second, the appearance of current peaks and negative differential resistance (NDR) at forward bias below 60 K. Since the NDR behavior is observed in the FE low temperature region, Mn substitution appears to induce a resonant state similar to a double barrier resonant tunneling diode \cite{Tsu}. A linear plot of the magnified \IV \ characteristics and the normalized conductance for \SRMO \ is shown in Fig. \ref{fig:Fig3_4.eps}(a) and (b), respectively. As the temperature is decreased, a current peak emerges at $\sim$ +0.25 V. Measurements taken under an applied magnetic field of 13 T revealed no significant difference in either the peak voltage or the conductance, indicating that this is a charge state, not a collective magnetic state.
\\\indent
The large decrease in the current can be quantitatively addressed by analyzing the temperature dependence of the junction ideality factor  which indicates the degree of spatial inhomogeneity of the Schottky barrier heights ($\Phi_{SB}$) \cite{Werner}. The assumptions of a Gaussian distribution and voltage dependence of the $\Phi_{SB}$ give the following relationship, which agrees with many experimental results\cite{Werner}:
\begin{equation}\label{eq:SBH_inhomo}
\frac{1}{n(T)}-1= \rho_2+\frac{\rho_3}{2kT/q}.
\end{equation}
Here, $n$ is the ideality factor, $k$ the Boltzmann constant, and $q$ the electronic charge. ${\rho}_2$ is a measure of the sensitivity of $\Phi_{SB}$ to the applied voltage and ${\rho}_3$ corresponds to the standard deviation in $\Phi_{SB}$. The present results plotted following  Eq. (\ref{eq:SBH_inhomo}) are shown in 
Fig. \ref{fig:Fig2a_4.eps}(b). The obtained values (${\rho}_2$, ${\rho}_3$) are (-0.57, 5.31 meV) for SrRuO$_3$ and (-0.30, 9.08 meV) for \SRMO. The increase in $\rho_3$ by almost a factor of two for the Mn-doped junction indicates that the presence of Mn ions increases potential fluctuations at the interface.
\subsection{\label{sec:modulation}Interface modulation}
\begin{figure}[t]
  \begin{center}
    \includegraphics[clip]{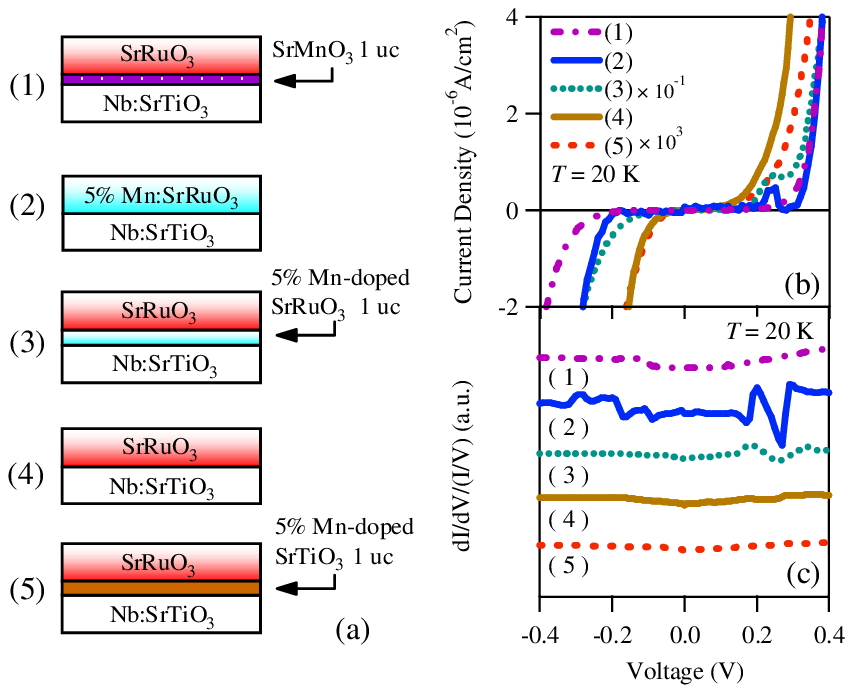}
  \end{center}
  \caption{(Color online) (a) Schematic diagram of the artificial structures grown to systematically vary the distribution of Mn ions with respect to the interface. (1) \SRO/[\SMO]$_{1 uc}$/\NSTO, (2) \SRMO/\NSTO, (3) \SRO/ [\SRMO]$_{1 uc}$/\NSTO, (4) \SRO/\NSTO, (5) \SRO/[\STMO]$_{1 uc}$/\NSTO. (b) \IV \ characteristics and (c) the normalized conductance of the structures illustrated in (a). NDR is observed only in structures (2) and (3).}
  \label{fig:Fig4_4.eps}
\end{figure}

To verify the role of Mn doping on the NDR, we fabricated modulated heterointerfaces where the Mn concentration was varied across the interface as shown in Fig. \ref{fig:Fig4_4.eps}(a). In the five structures, the location of the Mn ions is systematically varied with respect to the interface. Structures with single unit cell (uc) features were deposited while monitoring the reflection high-energy electron diffraction (RHEED) intensity. A two stage differential pumping setup enabled RHEED observations at high pressure.
\\\indent
In structure (1), one unit cell of \SMO \ is deposited on the \NSTO \ substrate at $P_{O2}$ = 0.1 Torr, followed by a thick layer of \SRO. The reduced current density in the whole voltage range at 20 K can be attributed to the high resistivity of \SMO \ at low temperatures, and as is evident from Fig. \ref{fig:Fig4_4.eps}(b), no NDR was observed.
In structure (2) the surface of the \NSTO \ is fully covered with \SRMO \ as discussed above. In structure (3), one unit cell of \SRMO, grown at $P_{O2}$ = 0.1 Torr, is inserted between a thick layer of \SRO \ and the \NSTO \ substrate. The basic features of the \IV \ characteristics are the same as in \SRO/\NSTO \ but with the presence of a current peak, implying that the origin of the observed NDR is not caused by the modification of the band alignment but the modification of the local interface structure. Structure (4) is the case of \SRO/\NSTO, which is shown again for comparison. The extreme case, simulating a situation where the Mn ions are embedded inside Nb:SrTiO$_3$, is demonstrated in structure (5). Here, a single unit cell of \STMO \ was deposited on \NSTO \ at $P_{O2}$ = 1.0 $\times$ 10$^{-5}$ Torr, followed by the deposition of SrRuO$_3$. From investigations of the leakage current at metal/Mn-doped \STO \ contacts, Mn-doped SrTiO$_3$ is known to be insulating \cite{Hofman, Morito}. In this case also no NDR was observed.
\\\indent
The above investigation has given evidence that the appearance of the low temperature NDR is associated with the presence of a low concentration of Mn ions on the metal side, ruling out the possibility of Mn, embedded in the substrate by diffusion or implantation acting as ionic potentials within the depletion region \cite{Modesti, caro}. The absence of the current peak in structure (1) is ascribed to the formation of a  \SMO \ \textit{band} rather than isolated atomic states at the interface.
\subsection{\label{sec:eels}Electron microscopy and spectroscopy}
\begin{figure}[t]
  \begin{center}
    \includegraphics[clip]{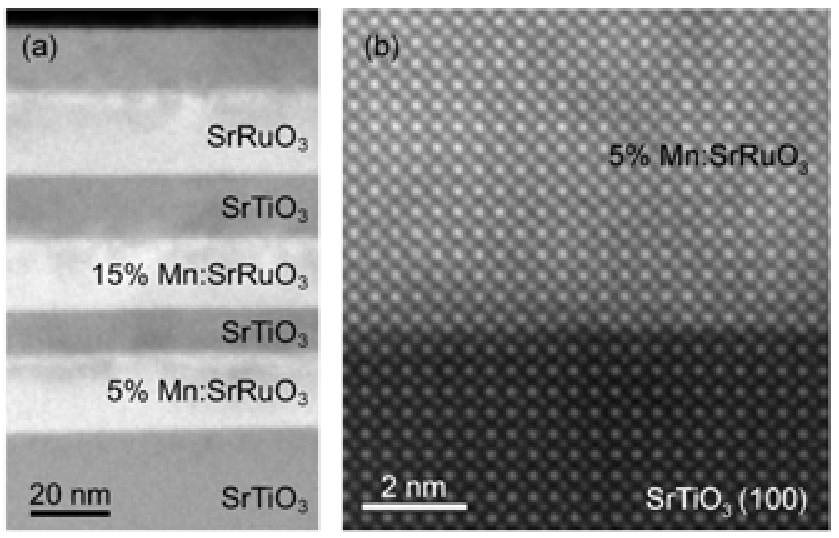}
  \end{center}
  \caption{Annular dark field STEM images of a (\SRMOx /\STO) \ multilayer. (a) Overview of the structure. (b) High magnification image of the abrupt \SRMO /\STO \ interface.}
  \label{fig:Fig5N_1.eps}
\end{figure}
\begin{figure}[t]
  \begin{center}
    \scalebox{0.7}{\includegraphics[clip]{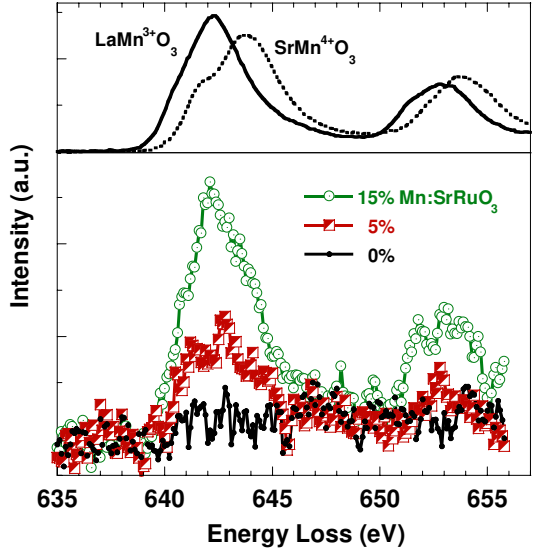}}
  \end{center}
  \caption{(Color online) Mn-L$_{2,3}$ electron energy loss spectra from the three Mn doped \SRO \ layers imaged in Fig. \ref{fig:Fig5N_1.eps}(a). For comparison Mn$^{3+}$ and Mn$^{4+}$ reference spectra are shown.}
  \label{fig:Fig6N_1.eps}
\end{figure}

The microscopic structure of the ruthanate/titanate interface was analyzed using scanning transmission electron microscopy (STEM) performed in a 200 kV FEI Tecnai F20 STEM. A superlattice of \SRO \ (20 nm)/\STO \ (10 nm)/SrRu$_{0.85}$Mn$_{0.15}$O$_3$ (20 nm)/\STO \ (10 nm)/\SRMO \ (20 nm) was fabricated on a \STO(100) substrate (Fig. \ref{fig:Fig5N_1.eps}(a)). 
An annular dark field (ADF) STEM image of the interface between the substrate and the first ruthanate layer, \SRMO, is shown in Fig. \ref{fig:Fig5N_1.eps}(b), demonstrating that the interface is chemically abrupt, with no obvious defects or dislocations. All Schottky junctions analyzed in this work were prepared under similar conditions as the test structure above, suggesting that the interfaces in these junctions are of comparable quality.   
In order to understand the local bonding of the Mn dopants in \SRMOx, electron energy loss spectroscopy (EELS) was performed. From the simple picture of Mn for Ru substitution in \SRMO, a Mn valence state of $4+$ is expected, however, Mn$^{3+}$ can also be present due to charge disproportionation in the form of Mn$^{4+}$ + Ru$^{4+}$ $\rightleftharpoons$ Mn$^{3+}$ + Ru$^{5+}$ (Ref. \onlinecite{Sahu}). In Fig. \ref{fig:Fig6N_1.eps}, the Mn-L$_{2,3}$ spectra for the three ruthanate layers are compared to Mn$^{3+}$ and  Mn$^{4+}$ reference spectra obtained from bulk LaMnO$_3$ and SrMnO$_3$, respectively \cite{target}. From the peak position of the Mn-L$_3$ edge a Mn valence of 3 $\pm$ 0.3 in the Mn doped ruthanate layers can be inferred. 

\section{\label{sec:discussion}discussion}
\begin{figure}[t]
  \begin{center}
    \includegraphics[clip]{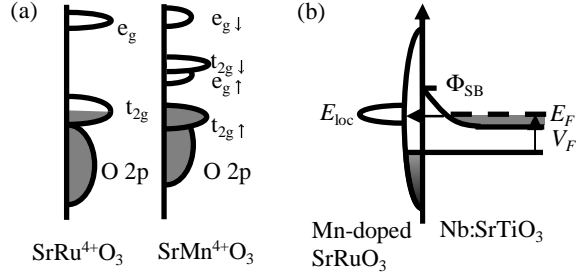}
  \end{center}
  \caption{(a) A schematic diagram of the density of states for \SRO \ and \SMO. (b) Interface band diagram illustrating the localized state on the metal side of the interface. $V_F$ is the applied forward bias voltage, $E_F$ is the Fermi level, $\Phi_{SB}$ the Schottky barrier height, and $E_{loc}$ is the localized state at the interface.}
  \label{fig:Fig7_4.eps}
\end{figure}

From the above studies, we have found that impurity doping on the metal side can strongly influence the electronic structure at the interface. We consider the relation between Mn doping in \SRO \ and the NDR by examining the electronic structures of the two parent compounds \SRO \ and \SMO. In \SRO, the crystal field splitting $10Dq$ of $\sim$ 3 eV is obtained from optical spectroscopy measurements \cite{Lee_b}, and the oxygen 2$p$ band lies below the Ru $t_{2g}$ band where the Fermi level is located as shown in Fig. \ref{fig:Fig7_4.eps}(a). We base our discussion on \SMO \ with a cubic G-type antiferromagnetic ground state for which the theoretically calculated density of states are schematically shown in Fig. \ref{fig:Fig7_4.eps}(a). The top of the valence band overlaps with the oxygen 2$p$ band and the Mn $e_g$ bands are located $\sim$ 0.3 eV above the Fermi level \cite{Sondena}. Within a rigid band picture, the Ru to Mn substitution modifies the empty states of \SRO \ arising from the Mn $e_g$ states. Furthermore, the doping independent lattice constant favors Mn ions to be localized owing to their smaller cation radius. 
\\\indent
The observation of the current peak at \textit{forward} bias and below 1.47 eV, the $\Phi_{SB}$ of \SRO/\NSTO(100) \cite{Hikita}, further support the above band structure consideration. Only when the localized states ($E_{loc}$) lie between the Fermi level ($E_F$) and $\Phi_{SB}$, can NDR be observed under \textit{forward} bias, as illustrated in Fig. \ref{fig:Fig7_4.eps}(b). In other cases, either the current will be dominated by the large FE current, or the localized states are buried in the large density of states of the host. We conclude that the NDR observed at the \SRMO/\NSTO \ interface is caused by the localized Mn $e_g$ states inside \SRMO \ acting as the resonant state.
\section{\label{sec:conclusion}conclusions}
We have applied the highly sensitive response of metallic perovskites to impurities in a Schottky junction to demonstrate that the interface electronic structure can be strongly modified by introduction of substitutional impurities on the \textit{metal} side. The change in the junction characteristics induced by impurity doping is unexpectedly large considering the small differences in the in-plane bulk transport properties. In addition to the enhanced spatial fluctuation of the interface potential, the dispersed Mn ions act as resonant states giving rise to NDR in the \IV \ characteristics. 
The poor electrostatic screening generally present in metallic perovskites and the close similarity in the chemical bonding between metallic and semiconducting perovskites can result in robust interface states exhibiting the character of the impurity element. Further studies may lead to designing resonant tunneling structures by simple doping of a metallic host, or as a prototypical structure for spectroscopic investigation of screening, localization, and metal-insulator transitions in strongly correlated electron systems.
\begin{acknowledgments}
This work was supported by the TEPCO Research Foundation and a Grant-in-Aid for Scientific Research on Priority Areas. Y.H. acknowledges support from the Grant-in-Aid 21st Century COE Program at the University of Tokyo. The work at Cornell University was supported under the ONR EMMA MURI monitored by Colin Wood. L. F. K. acknowledges financial support by Applied Materials.
\end{acknowledgments}

\appendix
\newpage 

\begin{thebibliography}{29}
\bibitem[{tar()}]{target}
\bibinfo{note}{LaMnO$_3$ bulk specimen was supplied by Toshima Manufacturing
  Co., Ltd. SrMnO$_3$ specimen was prepared by sintering SrCO$_3$ and MnO$_2$
  in molar ratio at 1300 $^\circ$C for 24 hours repeatedly with interval
  grinding.}
\end{thebibliography}

\end{document}